\documentclass[12pt]{article}
\usepackage{amsmath}
\usepackage{bm}
\usepackage{cite}
\usepackage{mathrsfs}
\usepackage{slashed}
\usepackage{graphicx}
\usepackage{hyperref}
\usepackage{verbatim}
\usepackage{color}
\usepackage{enumerate}
\usepackage{authblk}
\usepackage[hang,flushmargin]{footmisc}
\usepackage{lipsum}
\usepackage[font=footnotesize]{caption}
\usepackage{soul}
\usepackage[utf8]{} 
\usepackage{empheq}

\usepackage{amssymb,amsmath,amsfonts}
\usepackage[makeroom]{cancel}
\usepackage[normalem]{ulem}
\usepackage{hyperref}
\usepackage{color}

\newcommand{\diff}{\mathrm{d}}

\numberwithin{equation}{section}

\definecolor{blue-violet}{rgb}{0.54, 0.17, 0.89}
\definecolor{PineGreen}{cmyk}{0.92, 0, 0.59, 0.25}
\definecolor{OliveGreen}{cmyk}{0.64, 0, 0.95, 0.40}
\definecolor{RawSienna}{cmyk}{0, 0.72, 1, 0.45}
\definecolor{Gray}{cmyk}{0, 0, 0, 0.50}
\definecolor{MidnightBlue}{cmyk}{0.98, 0.13, 0, 0.43}
\definecolor{Orange}{cmyk}{0, 0.61, 0.87, 0}
\definecolor{LimeGreen}{cmyk}{0.50, 0, 1, 0}
\definecolor{Green}{cmyk}{1, 0, 1, 0}

\renewcommand{\d}{\partial}
\renewcommand{\tilde}{\widetilde}

\def\be{\begin{eqnarray}}
\def\ee{\end{eqnarray}}
\def\beann{\begin{eqnarray*}}
\def\eeann{\end{eqnarray*}}
\def\beq{\begin{equation}}
\def\eeq{\end{equation}}
\def\ba{\begin{array}}
\def\ea{\end{array}}
\def\ben{\begin{enumerate}}
\def\een{\end{enumerate}}
\def\bea{\begin{eqnarray}}
\def\eea{\end{eqnarray}}

\usepackage{atbegshi,picture}
\usepackage{lipsum}

\AtBeginShipoutNext{\AtBeginShipoutUpperLeft{  \put(\dimexpr\paperwidth-1cm\relax,-1.5cm){\makebox[0pt][r]{PUCV-UAI-23/03}}}}

\def\cH{\mathcal{H}}

\def\cL{\mathcal{L}}

\def\cN{\mathcal{N}}

\def\dD{\mathscr{D}}

\def\dI{\mathscr{I}}

\def\dL{\mathscr{L}}

\def\dP{\mathscr{P}}

\def\dU{\mathscr{U}}

\begin{document}

\title{\vspace{-70pt} \Large{\sc Kac-Moody symmetry in \\the light front of gauge theories} \vspace{10pt}}
\author[a]{\normalsize{Hern\'an A. Gonz\'alez}\footnote{\href{mailto:hernan.gonzalez@uai.cl}{hernan.gonzalez@uai.cl}}}
\author[b]{\normalsize{Oriana Labrin}\footnote{\href{mailto:orianalabrinz@gmail.com}{orianalabrinz@gmail.com}}}
\author[b]{\normalsize{Olivera Miskovic}\footnote{\href{mailto:olivera.miskovic@pucv.cl}{olivera.miskovic@pucv.cl}}}

\affil[a]{\footnotesize\textit{Facultad de Artes Liberales, Universidad Adolfo Ib\'a\~nez, Diagonal Las Torres 2640, Santiago, Chile.}}
\affil[b]{\footnotesize\textit{Instituto de F\'isica, Pontificia Universidad Católica de Valparaíso, Avda.~Universidad 330, Curauma, Valparaíso, Chile.}}

\date{}

\maketitle

\begin{abstract}
We discuss the emergence of a new symmetry generator in a Hamiltonian realisation of four-dimensional gauge theories in the flat space foliated by retarded (advanced) time.  It generates an asymptotic symmetry that acts on the asymptotic fields in a way different from the usual large gauge transformations. The improved canonical generators, corresponding to gauge and asymptotic symmetries, form a classical Kac-Moody charge algebra with a non-trivial central extension. In particular, we describe the case of electromagnetism, where the charge algebra is the $\mathrm{U}(1)$ current algebra with a level proportional to the coupling constant of the theory, $\kappa=4\pi^2/e^2$. We construct bilinear generators yielding  Virasoro algebras on the null boundary. We also provide a non-Abelian generalization of the previous symmetries by analysing the evolution of Yang-Mills theory in Bondi coordinates. 
\end{abstract}

\newpage
\tableofcontents

\section{Introduction}

Symmetry is the guiding principle to unravel universal properties and dualities arising in quantum field theories. In this sense, asymptotic symmetries have served as a crucial tool to disclose these features. Going almost fifty years back, the work of Brown and Henneaux has shown that the Virasoro group plays an important part in the realization of holography on AdS$_{3}$ spaces \cite{Brown:1986nw}. More recently, we have witnessed the importance of the BMS asymptotic symmetry, displayed by gravity at null infinity \cite{Sachs:1962wk,Bondi:1962px}. This group is enriched with an infinite set of angle-dependent translations, making it difficult to isolate a unique global Lorentz subgroup\footnote{Very recently, a new set of relaxed boundary conditions at spatial infinity has been constructed in \cite{Fuentealba:2022xsz,Fuentealba:2023rvf,Fuentealba:2023huv} for gravity and electromagnetism. Making nonlinear redefinitions of the generators, it is shown that the asymptotic charge algebra is a direct sum of Poincaré and supertranslation modes. See also \cite{Javadinezhad:2022hhl,Capone:2022gme, Compere:2023qoa} for related discussions.}. Nowadays, these extra generators, called supertranslations, have been reinterpreted (see \cite{Strominger:2017zoo} for a thorough summary of these developments in gravity and gauge theory) as part of the universal organization of the leading infrared  pole exhibited by the Weinberg soft theorems \cite{Weinberg:1965nx}. 

The Hamiltonian formulation of field theories is a systematic method to consistently determine all the symmetries and degrees of freedom present in a given model with the prescription of the initial data on a Cauchy surface.
Evolution along null hypersurfaces are particularly useful in the study of gravitational waves and the behavior of fields at null infinity. However, using the retarded/advanced time as the parameter of the evolution, we find an obstruction that only a part of the degrees of freedom can be prescribed initially, so the Hamiltonian evolution becomes incomplete. One way of dealing with this problem has been presented in \cite{Nagarajan:1985xn} for scalar and Maxwell fields. Gluing together a spacelike and null surfaces,  it was possible to single out an initial data Cauchy surface.  

In this work, we will adopt a different approach to analyse a Hamiltonian evolution of a gauge theory in the null foliated Minkowski spacetime. In our settings, the light-front dynamics \cite{Dirac:1949cp} is analysed by looking at a spherically symmetric system evolving along retarded or advanced time. Contrary to \cite{Nagarajan:1985xn}, we investigate the consequences of prescribing initial data only on a null surface. We will see that one of the effects, arising under these conditions in an evolution of a system, consists  in appearance of a new constraint whose zero mode will give rise to an additional local symmetry.  

To shed more light on this problem let us perform a canonical analysis of the two-dimensional chiral boson theory \cite{Floreanini:1987as}. The Lagrangian of a real chiral scalar field $\Phi(t,\varphi)$ is given by 
\be
\label{chib}
\cL(\Phi)=\d_\varphi  \Phi \, \d_t \Phi - (\d_\varphi \Phi)^2\,.
\ee
Since it is linear in the velocities $\dot \Phi$,    the momentum definition $\pi_\Phi=\frac{\d\cL}{\d\dot{\Phi}}=\d_\varphi \Phi$ leads to a primary constraint, better analysed in terms of its Fourier modes
$
C_n=\oint \diff \varphi\, \mathrm{e}^{\mathrm{i}n \varphi} (\pi_\Phi-\d_\varphi \Phi)
$, $n \in \mathbb{Z}$. 
The corresponding Poisson bracket algebra reads
\be
\{C_n,C_m\}=4\pi \mathrm{i} \, m\delta_{n+m,0} \,.
\ee
We conclude that, except for $C_0$, all other $C_n$ are second class constraints. In fact, the first class generator $C_0$  produces a genuine  symmetry in \eqref{chib}, given by the finite transformation $\Phi(t,\varphi) \to \Phi(t,\varphi) + \theta(t)$.  
Similar aspect can be also observed in a non-Abelian generalization of this model, corresponding to either a multiplet of non-Abelian chiral bosons, or WZNW model. In both cases, the appearance of zero mode generators is related to the existence of centrally extended Kac-Moody algebras \cite{Salomonson:1988mk,Sazdovic:1995np}.

In fact, any second order Lagrangian written in the light-cone coordinates adopts a form similar to \eqref{chib}, that is, becomes a first order Lagrangian in the advanced/retarded time frame \cite{Alexandrov:2014rta}. Furthermore, in higher
dimensions, it is also known that the bulk gauge symmetries can become large transformations by imposing asymptotic conditions on the fields \cite{Benguria:1976in}. 

In this work we will address the question of the soft charges associated to zero modes analogous of $C_0$. With a suitable incorporation of asymptotic falloffs, the generators will be enhanced by a surface term representing  an additional global charge. Interestingly enough,  the algebra of all asymptotic charges corresponds to a generalization of a Kac-Moody algebra endowed with a non-trivial central extension. We will workout in detail the cases of electromagnetism and Yang-Mills theory.  Residual gauge transformations in light-cone electromagnetism in four dimensions have been recently discussed in \cite{Majumdar:2022fut}.

\section{Electromagnetism in the null frame}

\label{H procedure}

We analyse an evolution of a four-dimensional gauge theory in a reference frame associated to the radiation waves, that is, where the flat space-time has been foliated with null surface slices. In the spherical coordinates $(t,r,\theta ,\varphi )$, it means that the system is evolving in the retarded (advanced) time defined by
\begin{equation} 
u=t-\epsilon r\in \mathbb{R}\,,\qquad \epsilon \in [-1,1]\,,  \label{u}
\end{equation}%
where the parameter $\epsilon $ is  related to the inverse velocity of the
particle such that, for $\epsilon =\pm 1$, the coordinate $u$ becomes lightlike. In fact, the retarded time corresponds to $\epsilon =1$, the usual time (associated to the proper time of a massive particle) has $\epsilon =0$ and the advanced time is obtained when $\epsilon =-1$. To  understand better how the system reaches the limit $\epsilon \to \pm 1$, we choose the local coordinates $x^{\mu }=(u,r,\theta ,\varphi )$ for general $\epsilon $. The Minkowski metric in the \emph{Bondi coordinate frame} reads
\begin{eqnarray}
\label{flat}
\diff s^{2} &=& \mathfrak{g}_{\mu\nu}(x)\,\diff x^\mu \diff x^\nu= -\diff t^{2}+\diff r^{2}+r^{2}\diff \Omega ^{2}  \notag \\
&=&-\diff u^{2}-2\epsilon\, \diff u \diff r+\left( 1-\epsilon ^{2}\right) \diff r^{2}+r^{2}\diff \Omega ^{2}\,,
\end{eqnarray}
where the line element of the unit 2-sphere is
\begin{equation}
\diff \Omega ^{2}=\gamma _{AB}(y )\,\diff y ^{A}\diff y ^{B}= \diff \theta^2+\sin^2\theta \diff \varphi^2\,.  \label{omega}
\end{equation}
In order to simplify later computations, it is convenient to introduce the complex coordinates $y^{A}=(z,\bar{z})$ corresponding to the stereographic projection of the sphere to the complex plane, as
\begin{equation}
z=\mathrm{e}^{\mathrm{i}\varphi } \cot \frac{\theta }{2}\,. \label{stereographic}
\end{equation}
The metric of the two-sphere becomes
\begin{equation}
\gamma _{AB}=\left(
\begin{array}{cc}
0 & \gamma _{z\bar{z}} \\
\gamma _{z\bar{z}} & 0
\end{array}
\right) ,\qquad \gamma _{z\bar{z}}=\frac{2}{(1+z\bar{z})^{2}}\,.
\end{equation}
In these coordinates, the Minkowski metric and the Jacobian have the form 
\begin{equation}
\mathfrak{g}_{\mu \nu }=\left( 
\begin{array}{ccc}
-1 & -\epsilon & 0 \\ 
-\epsilon & 1-\epsilon ^{2} & 0 \\ 
0 & 0 & r^{2}\gamma _{AB}
\end{array}
\right) ,\quad  \sqrt{\mathfrak{g}}=r^{2}\sqrt{\gamma }\,,\label{metric}
\end{equation}
where  $\mathfrak{g}$ and $\gamma$ denote the absolute values of the determinants $\mathfrak{g}_{\mu \nu}$ and $\gamma_{AB}$, respectively.  With this notation at hand, we write the electromagnetic action in the background metric $\mathfrak{g}_{\mu \nu}$ as 
\begin{equation}
I[A]=\int \diff ^{4}x\,\mathcal{L}(A,\partial A)=-\frac{1}{4 e^2}\int \diff ^{4}x\,\sqrt{\mathfrak{g}}\,F^{\mu \nu
}F_{\mu \nu }\,,  \label{action}
\end{equation}
where $A_\mu(x)$ is the Abelian gauge field, $F_{\mu \nu}=\partial_\mu A_\nu - \partial_\nu A_\mu$ is the Maxwell's electromagnetic tensor and $e$ is the coupling constant of the theory, the electron charge. 

\subsection{Hamiltonian formulation}

In order to pass to the Hamiltonian formalism, we define the canonical momentum, which is a tensor density, in the retarded time foliation as 
\begin{equation}
\pi ^{\mu }=\frac{\partial \mathcal{L}}{\partial \dot{A}_{\mu }}=-\frac{1}{e^2}\,\sqrt{\mathfrak{g}}\,F^{u\mu
}\,,
\end{equation}
where the velocity $\dot{A}_{\mu }=\partial _{u}A_{\mu }$ is defined along a timelike vector field  $\partial _{u}$ tangent to the coordinate $u$, which describes a direction of propagation.

In components $\pi ^{\mu }=(\pi ^{u},\pi ^{r},\pi ^{A})$, it reads 
\begin{eqnarray}
\pi ^{u} =0\,,  \qquad 
\pi ^{r} =\frac{r^{2}}{e^2}\,\sqrt{\gamma }\,F_{ur}\,, \qquad
\pi ^{A} =-\frac{1}{e^2}\,\sqrt{\gamma }\gamma ^{AB}\left[
(\epsilon ^{2}-1)F_{uB}-\epsilon F_{rB}\right] \,,   \label{momenta}
\end{eqnarray}
making it evident that the cases $\epsilon ^{2}=1$ and $\epsilon^{2}<1$ lead to different Hamiltonian structures. Indeed, when $\epsilon^{2}<1$, the velocity $\dot{A}_{A}$ can be expressed in terms of $\pi ^{A}$. However, when $\epsilon ^{2}=1$, this is not possible and the relation involving $\pi ^{A}$ becomes a constraint. Thus, the limit $\epsilon ^{2}\rightarrow 1$ is discontinuous and the Hamiltonian dynamics of a theory on the light front cannot be obtained as a limiting case of its Hamiltonian description studied in the frame of a massive particle.

An appearance of a new constraint is not only a feature of gauge theories, but of any dynamical theory defined on the light front, because it is a consequence of the fact that the action written in the light-cone coordinates becomes a first order one \cite{Steinhardt:1979it,Alexandrov:2014rta}.
A difference between Hamiltonian approaches and symmetry analysis based on different choices of time foliation, $\epsilon=0$ or $\epsilon=\pm 1$, has been discussed in \cite{Blagojevic:1993fp}. 
To explore this last aspect of gauge theories, we will focus on the Bondi reference frame, namely, take
\begin{equation}
\epsilon ^{2}=1\,. \label{epsilon^2=1}
\end{equation}
From now on, we assume that the above condition is always satisfied.

Geometrically, a radial wave located at $(t,r)$ propagates along the null coordinate $u$, such that we study the evolution  across different  $u=const$ surfaces.\footnote{The null coordinates $t-\epsilon r=u$ and $t+\epsilon r=u+2\epsilon r$ parameterize the cone $\diff s^2=0$, describing physically equivalent geometries that correspond to the opposite signs of $\epsilon $.} Thus, in absence of massive particles, the future lightlike infinity $\dI^+$, parametrized by $(u|_{\epsilon=+1},y^A)$, and the past lightlike infinity $\dI^-$, parametrized by $(u|_{\epsilon=-1},y^A)$, behave as Cauchy surfaces \cite{Hawking:2016msc}.\footnote{For the Penrose diagram showing a photon propagation in the Minkowski space see, for example, ref.~\cite{Strominger:2017zoo}.} In this coordinate frame, the photons moving in the future cone $\epsilon =+1$ and the past cone  $\epsilon =-1$ are described by two different Hamiltonians. They satisfy the antipodal matching conditions $\mathcal{H}_T|_{\dI^-_+}=\mathcal{H}_T|_{\dI^+_-}$  that identify their values in the antipodal points on the sphere, one at the future of $\dI^-$ and another at the past of $\dI^+$  \cite{Kapec:2015ena,Campiglia:2015qka}.

When \eqref{epsilon^2=1} is fulfilled, it can be seen from eqs.~\eqref{momenta} that  there are the following \emph{primary constraints}, 
\begin{equation}
\pi ^{u}\approx 0\,,\qquad \chi ^{A}\equiv \epsilon \pi ^{A}-\frac{1}{e^2} \,\sqrt{\gamma }
\gamma ^{AB}F_{rB}\approx 0\,,  \label{primary}
\end{equation}
where the weak equality `$\approx$' is used for the constraints to point out that these functions are zero, but their Poisson brackets are not zero in general. 

Let us denote the spacelike indices by $x^{i}=(r,y ^{A})$, such that the coordinate labels are summarized as
\begin{equation}
x^{\mu }=(u,r,y^{A})=(u,x^{i})\,,\qquad x^{i}=(r,y^{A})\,,\qquad
y^{A}=(z,\bar z)\,.
\end{equation}
Now we can compute the \emph{canonical Hamiltonian}, $\mathcal{H}
_{C}=\pi^{\mu }\dot{A}_{\mu }-\mathcal{L}$, as 
\begin{equation}
\mathcal{H}_{C}=\frac{e^2\left( \pi ^{r}\right) ^{2}}{2 r^{2}\sqrt{\gamma }}+
\frac{e^2 \tilde{\pi}_{A}\pi^{A}}{2\sqrt{\gamma }}+\frac{\sqrt{\gamma }}{4e^2 r^{2}}\,%
\tilde{F}^{AB}F_{AB}-A_{u}\partial _{i}\pi ^{i}\,,  \label{HC}
\end{equation}
where we neglected a total divergence $\partial_i(A_u\pi^i)$ that will become a boundary term after integrating in the spatial coordinates. We will take care of the boundary terms in Subsection \ref{U(1)}. The metric $\gamma _{AB}$ and its inverse $\gamma ^{AB}$ raise and
lower the indices on the 2-sphere. Furthermore, we use  the notation
\begin{equation}
    \tilde{V}^A=\gamma^{AB}V_B\,,
\end{equation}
when raising or lowering the indices using the metric of the 2-sphere\footnote{In particular, $\tilde{\pi}_{A}=\gamma _{AB}\pi ^{B}$ and $\tilde{F}^{AB}=\gamma ^{AC}\gamma
^{BD}F_{CD}$. Therefore, $\pi _{\mu }=g_{\mu \nu }\pi ^{\nu }$ and $F^{\mu
\nu }=g^{\mu \alpha }g^{\nu \beta }F_{\alpha \beta }$ projected to the
sphere are related to $\tilde{\pi}_{A}$ and $\tilde{F}^{AB}$ as
$
\pi _{A}=r^{2}\tilde{\pi}_{A}\,,F^{AB}=r^{-4}\,\tilde{F}^{AB}\,.
$}.
Following the Dirac procedure \cite{dirac2001lectures}, we define the \emph{total Hamiltonian} that includes all primary constraints along the indefinite multipliers $\lambda _{u}(x)$ and $\lambda _{A}(x)$ associated to the constraints \eqref{primary}, such that
\begin{equation}
\mathcal{H}_{T}=\mathcal{H}_{C}+\lambda _{u}\pi ^{u}+\lambda _{A}\,\chi ^{A}\,.
\label{HT}
\end{equation}
Note that the Lagrangian $\mathcal{L}$ in \eqref{action} is a scalar density and the momenta $\pi ^{\mu }$ are vector densities  because they both include the Jacobian $\sqrt{\mathfrak{g}}$ in the definition, which implies that the Hamiltonian density $\mathcal{H}$ also includes the Jacobian. Therefore, the Hamiltonian is defined without  $\sqrt{\mathfrak{g}}$ in the volume element,
\begin{equation}
H_{T}=\int \diff ^{3}x\,\mathcal{H}_{T}\,. \label{intHt}
\end{equation}
In that way, an evolution of a function $\psi(A(x),\pi(x),\lambda(x))=\psi(x)$ of canonical variables and multipliers is given by
\begin{equation}
\dot{\psi}(x)\approx\{\psi (x),H_{T}\}=\int \diff^{3}x'\,\{\psi (x),\mathcal{H}_{T}(x')\}\,.
\end{equation}
The canonical Poisson brackets (PB) have the form 
\begin{equation}
\left\{ A_{\mu }(x),\pi ^{\nu }(x')\right\} =\delta _{\mu }^{\nu }\,\delta ^{(3)}(x-x')\,,
\end{equation}
where $u=u'$ is always assumed. We will not  write explicitly the dependence of the fields on the retarded (advanced) time variable $u$.

To conclude, the Hamilton's equations of the generalized coordinates are 
\begin{eqnarray}
\dot{A}_{u} \approx \lambda _{u}\,,  \quad 
\dot{A}_{r} \approx\frac{e^2 \pi^{r}}{r^{2}\sqrt{\gamma }}+\partial _{r}A_u \,, \quad
\dot{A}_{A}\approx\frac{e^2\tilde\pi _{A}}{\sqrt{\gamma }}+ \partial _{A}A_u +\epsilon\lambda _{A}\,,  \label{dot A}
\end{eqnarray}
where $\dot{A}_{A}$ and $\dot{A}_{u}$ are not expressible in terms of momenta, since their evolution is given by the multipliers. The Hamilton's equations of the canonical momenta have the form
\begin{eqnarray}
\dot{\pi}^{u}\approx \partial _{i}\pi ^{i}\approx 0\,,  \quad
\dot{\pi}^{r} \approx\frac{\sqrt{\gamma }}{e^2}\,\nabla_{A}\tilde\lambda^{A}\,,  \quad
\dot{\pi}^{A} \approx-\frac{\sqrt{\gamma }}{e^2}\left( \frac{1}{r^{2}}\,\nabla_{B}\tilde{F}^{AB}+ \partial _{r}\tilde\lambda^{A}\right) .  \label{dot P}
\end{eqnarray}

\subsubsection*{Consistency conditions}

For consistency of the theory, the constraint surface has to be conserved during its evolution. To explore it, let us first compute the auxiliary PBs 
\begin{equation}
\left\{ \chi ^{A}(x),\chi ^{B}(x')\right\} =\Omega
^{AB}(x,x')\equiv -\frac{2\epsilon}{e^2} \sqrt{\gamma }\gamma ^{AB}\partial
_{r}\delta ^{(3)}(x-x')\,,   \label{Omega}
\end{equation}
where $\Omega ^{AB}(x,x')$ is the symplectic matrix, whose role
will be analysed later. Then,  we find that the consistency condition of the first primary constraint leads to a  secondary constraint,\footnote{We are using a shorthand notation for the functions in PBs: $\psi\equiv \psi(x)$ and $\psi'\equiv \psi(x')$.}
\begin{equation}
\dot{\pi}^{u}=\int \diff^{3}x'\left\{ \pi ^{u},\mathcal{H}_{T}^{\prime
}\right\} =0\quad \Rightarrow \quad \chi =\partial _{i}\pi ^{i}\approx 0\,, \label{secondary}
\end{equation}
where the consistency of the secondary constraint does not give anything new, 
$\dot{\chi} =0$.  Similarly, the consistency of the primary constraint $\dot{\chi}^A =0$ leads to a differential equation for its indefinite multiplier, 
\begin{align}
\d_{r}\lambda _{A} & =-\frac{\epsilon e^2 }{2 \sqrt{\gamma }} \, \d_r \tilde\pi_{A}-\frac{1}{2 r^{2}}\,
\nabla^{B}F_{AB}+\frac{\epsilon e^2}{ 2r^{2}}\,\partial _{A}\left( \frac{\pi ^{r}}{\sqrt{\gamma }}\right)  
,  \label{lambdaEq}
\end{align}
 where $\nabla=\nabla(\gamma )$ is the covariant derivative on the sphere. Therefore, the
multipliers $\lambda_{A}$ can be solved from \eqref{lambdaEq} as
\begin{equation}
\lambda_{A}=\Lambda_{A}- \frac{\epsilon e^2}{2\sqrt{\gamma }}\,\tilde\pi _{A}+\int\limits^{r}
\frac{\diff r}{2r^2}\left[ \epsilon e^2\,\partial _{A}\left( \frac{\pi^{r}}{\sqrt{
\gamma }}\right) -\,\nabla^{B}F_{AB}\right] \equiv \Lambda_A + \bar \lambda_A\,
,  \label{bar}
\end{equation} 
where $\bar \lambda_A$ is a determined part of the multiplier and $\Lambda _{A}(y)$ are the $r$-independent functions that remain
arbitrary. Since each arbitrary multiplier is associated to a first class
constraint, the existence of $\Lambda _{A}$ suggests that there exist $r$-independent first class constraints. We will see later that this is indeed the case.\medskip

In sum, our constraints are:
\begin{equation}
\fbox{$
\begin{array}{ll}
\text{Primary constraints:} & \pi ^{u}\,,\quad \chi ^{A}=\epsilon \pi ^{A}-\frac{1}{e^2} 
\sqrt{\gamma }\gamma ^{AB}F_{rB}\,, \\ 
\text{Secondary constraint:} & \chi =\partial _{i}\pi ^{i}\,.
\end{array}$}
\end{equation}

We could also introduce the extended Hamiltonian, $\mathcal{H}_{E}=\mathcal{H}_{T}+
\lambda_\chi\chi $, that contains all first class constraints present in the theory, and a new multiplier $\lambda_\chi(x)$,  that describes the most general evolution of the system. However, we will omit this step, as our results do not depend on it, and the gauge field $A_u$   plays a role of the  multiplier for the constraint $\chi$ in the canonical Hamiltonian \eqref{HC}.

\subsubsection*{First and second class constraints}

We have to investigate next the class of the above constraints. As well-known, the first class constraints have PBs weakly equal to zero with all other ones, and they are related to existence of local symmetries in a theory. On the other hand, second class constraints have invertible PBs, and they only remove redundant fields from the theory.

Direct computation shows that the only non-vanishing PBs among all the constraints are the ones given by eq.~\eqref{Omega}. So, $\pi ^{u}$ and $\chi $ are first class. In fact, the symmetry associated to them is the local $\mathrm{U}(1)$, as $\chi=0$  corresponds to the differential Gauss law. Besides, the class of  $\chi ^{A}$ depends on the symplectic matrix $\Omega^{AB}$ given by \eqref{Omega}. Seen as a $2 \times 2$ matrix, $\Omega^{AB}$ is invertible because the metric $\gamma ^{AB}$ is invertible. Therefore, the constraints $\chi ^{A}$ are second class, that is, they do not generate any local symmetry, but they just eliminate non-physical components of the canonical fields. Indeed, this approach has been adopted in refs.~\cite{Goldberg:1991pb,Majumdar:2022fut}, where non-dynamical variables resulting from second class constraints were eliminated from the phase space by solving $\chi ^{A}=0$.

Nonetheless, the inverse of $\Omega_{AB}$ is not unique. A reason is that, strictly speaking, the symplectic matrix $\Omega^{AB}$ is infinite-dimensional, as it also depends on continuous variables. This still allows for a possibility to have first class constraints in the continuous sector of the constraints. To analyse it, we recall that the invertibility of any matrix is equivalent to  the absence of its zero modes. However, in our case it can be shown that the symplectic matrix $\Omega ^{AB}(x,x')$ has $r$-independent zero modes $V_{A}$, which are a solution of the corresponding zero mode equation 
\begin{equation}
0=\int \diff^{3}x'\,\Omega ^{AB}(x,x')V_{B}(x')=-\frac{2\epsilon}{e^2} \, \sqrt{\gamma }\gamma ^{AB}\partial _{r}V_{B}\quad \Rightarrow
\quad V_{B}=V_{B}(y)\,.
\label{zeroeq}
\end{equation}
Therefore, the invertible part of $\Omega ^{AB}$ corresponds to the second class constraints and the zero modes correspond to the first class constraints, related to the undetermined multipliers $\Lambda _{A}(y)$. From the point of view of the bulk, the
symmetries generated by them are not a local symmetry of a theory, but an  infinite number of global symmetries. 

One way to separate explicitly first class constraint from the rest of constraints in $\chi ^{A} $ is to expand all the fields asymptotically in the vicinity of the boundary,  located at the constant radius $r\to \infty$. Then the first class constraint is the $r$-independent part of $\chi^{A}(x)$ in the expansion, which can be called $\chi _{(0)}^{A}(y)$. 

It can be checked straightforwardly that all PBs of the constraints $\pi ^{u}$ and $\chi $ are zero, and that they also commute with $\chi ^{A}$. Moreover, it  can be shown that $\chi _{(0)}^{A}$ are indeed first class if we compute their PBs (given by the distribution $\Omega ^{AB}$) under an integral, with an arbitrary test function $\psi (x)$. Then they give zero only on the space of test  functions that satisfy $\partial _{r}\psi =0$, confirming that only the $r$-independent part of $\chi ^{A}$ are commuting. Later, we will also  prove that only $\chi _{(0)}^{A}$  among  all $\chi ^{A}$ generate the symmetry transformations of the canonical fields and  multipliers.

\subsection{\texorpdfstring{$\mathrm{U}(1)$}{U(1)} local symmetry and  asymptotic symmetries} \label{U(1)}

We will assume the \emph{standard asymptotic conditions} in electromagnetism in Bondi coordinates \cite{Strominger:2013lka}. In the Hamiltonian formulation, they are given by
\begin{equation}
\begin{array}{llllll}
A_{u} & =\mathcal{O}\left( r^{-1}\right) \,,\quad  & A_{r} & =\mathcal{O}\left( r^{-2}\right) \,,\quad  & A_{A} & =\mathcal{O}(r^{0})\,,\medskip  \\
\pi ^{u} & =0\,, & \pi ^{r} & =\mathcal{O}\left( r^{0}\right) \,, & \pi ^{A}
& =\mathcal{O}\left( r^{-2}\right) \,. \label{standard}
\end{array}
\end{equation}
Furthermore, the multipliers behave as
\begin{equation}
\lambda _{u}=\mathcal{O}\left(r^{-1}\right) \,,\qquad \Lambda_{A}=\mathcal{O}\left( r^0\right) \,,\label{StandardMultipliers}
\end{equation}
where also $\bar\lambda_A =\mathcal{O}\left(r^{-1}\right)$. This is a consequence of the Hamilton's equations \eqref{dot A} and \eqref{dot P}. 

In general, not all asymptotic conditions are invariant under Poincaré transformations. For instance, when studying the Hamiltonian dynamics at spatial infinity of electromagnetism \cite{Henneaux:2018gfi} and Yang-Mills theory \cite{Tanzi:2020fmt}, one needs to supplement the phase space with parity conditions on the fields, to ensure that the evolution of the theory preserves the canonical form of Lorentz boosts.  In  the present case, though,  it has been verified in \cite{Bunster:2018yjr} that the conditions \eqref{standard} are invariant under Poincaré group using a time-like foliation that approaches to the null infinity after a specific limit is taken. 

In what follows, we will construct and discuss the generators of symmetries. As we already saw, each first class constraint has an associated  indefinite multiplier. In addition, each of them will generate a local symmetry, with a corresponding local parameter and an associated conserved  charge. This is summarized in the following table:
\begin{equation}
\fbox{$
\begin{array}{ccccc}
\text{First class constraints} & \text{Multipliers} & \text{Parameters} & 
\text{Generators} & \text{Charges} \\ 
\pi ^{u},~\chi =\partial _{i}\pi ^{i} & \lambda_{u},~-A_u
& \theta _{u},~\theta & G[\theta ] & Q[\theta ] \\ 
\chi_{(0)}^{A}=(\epsilon\pi^{A}-\frac{1}{e^2}\sqrt{\gamma }\gamma^{AB}F_{rB})_{(0)} & \Lambda
_{A}(y) & \eta _{A}(y ) & S[\eta ] & Q_s[\eta ]
\end{array}
$} \label{summary}
\end{equation}

Thus, let us define two kind of smeared generators associated to the above sets of first class constraints, 
\beq
\begin{split}
G[\theta ] &=\int \diff ^{3}x\,\left( \theta \,\partial _{i}\pi ^{i}+\theta_{u}\pi ^{u}\right) \,, \\
S[\eta ] &=\int \diff ^{3}x \, \eta _{A} \chi^{A} \,.   \label{G,S}
\end{split}
\eeq
The generator $G$ will give rise to the usual $\mathrm{U}(1)$ charge, namely the electric charge for the parameter $\theta$ that goes to a constant value when $r \to \infty$, and $S$ will give new asymptotic charges related to so-called magnetic large gauge transformations \cite{Campiglia:2016hvg}. 

Let us remind that second class constraints do not form a part of symmetry generators in general, but in our expression for $S$ they do. This is why we would like to ensure that there is only an effect of first class constraints contained in $S$.

Mathematically, a reason for including all coefficients $\chi _{(n)}^{A}$ ($n\geq 0$) in the Taylor expansion of $\chi ^{A}$ in $1/r$ is that this expansion is not invertible when $r$ is a non-compact coordinate, i.e., the
basis $\{r^{n}|n\in \mathbb{N}_{0}\}$ is not complete. Thus, a  way to ensure that the second class constraints $\chi _{(n)}^{A}$ ($n\geq 1$) do not contribute to the transformation law of the fields, is to assume that the parameters $\eta _{A}$ tend very fast to  radially-independent functions at the boundary, such that we can take $\eta _{A}=\eta _{A}(y)$. We show below that this singles out only a contribution of the first class
constraint $\chi _{(0)}^{A}$ in the transformation law of the fields.
Indeed, let us compute the transformations $\delta _{\theta }\equiv \{\;\, , G[\theta]\}$ and $\delta _{\eta }\equiv \{\; \,, S[\eta]\}$ generated by the  functionals \eqref{G,S}, assuming that integration by parts does not produce any boundary terms. We find 
\begin{eqnarray}
\label{PBth}
\delta _{\theta }A_u =\theta _{u}\,, \qquad
\delta _{\theta }A_i =-\partial _{i}\theta \,, \qquad
\delta _{\theta }\pi ^{\mu } =0\,, 
\end{eqnarray}
where we recognize the usual $\mathrm{U}(1)$ gauge transformations in Hamiltonian formulation.\footnote{The transformation law acquires the usual Lorentz-covariant form $\delta_\theta A_\mu=-\partial_\mu \theta$ by means of the Castellani's method \cite{Castellani:1981us}, which requires  $\theta^u=-\dot \theta$ in the generator \eqref{G,S}.}

As regards the second generator, the transformations act only at the radially independent components, that is, at infinity,
\begin{eqnarray}
\label{PBeta}
\delta _{\eta }A_{\mu } =\epsilon\eta _{A}\,\delta _{\mu
}^{A}\,,  \qquad
\delta _{\eta }\pi ^{\mu } 
=\frac{1}{e^2}\, \delta _{r}^{\mu }\sqrt{\gamma }\,\nabla_{A}\tilde\eta^{A}\,. 
\end{eqnarray}
Indeed, because the parameters $\eta $ do not depend on the radial coordinate, the only coefficients in the expansion that transform non-trivially are the leading coefficients of the fields,
\begin{equation}
\delta _{\eta }A_{(0)\mu} =\epsilon\, \eta _{A}\,\delta^A_{\mu}\,,  \qquad
\delta _{\eta }\pi _{(0)}^{r} =\frac{1}{e^2}\,\sqrt{\gamma }\,\nabla_{A}\tilde\eta^{A}\,,
\label{boundary transformations} 
\end{equation}
while all others remain  invariant. 

It can be checked straightforwardly that the Hamilton's equations \eqref{dot A} and \eqref{dot P} are invariant under the gauge transformations \eqref{PBth}--\eqref{PBeta} if the multipliers transform as
\begin{eqnarray}
\delta_{\theta,\eta} \lambda _{u} =\dot{\theta}_{u}\,,  \qquad \delta_{\theta,\eta}A_u =-\dot{\theta}+\mathcal{O} (r^{-1})\,, \qquad
\delta_{\theta,\eta}\Lambda _{A} = \dot{\eta}_{A}+\mathcal{O} (r^{-1}) ,
\end{eqnarray}
where the $\mathcal{O} (r^{-1})$ terms depend only on $\eta_{A}$ and, in order to preserve the falloff given by \eqref{standard}, the local parameter $\theta$ does not depend on $u$ at the leading order. Thus,  the invariance under transformations with the parameter $\eta_{A} \neq 0$ is fulfilled only in the limit $r\to\infty$ and the symmetry is truly asymptotic.  

Note that the Hamilton's equation \eqref{dot A} for $A_A$ implies that $\Lambda _{A}=\epsilon \dot{A}_{A}+\mathcal{O}(r^{-1})$. The above transformation law is consistent with this relation, as it can be seen from  \eqref{boundary transformations}. Nonetheless, the relation between $\Lambda_A$ and $\dot A_A$ is valid only \emph{on-shell}.

Finally, if the fields behave asymptotically as in  eq.~\eqref{standard}, the transformation laws \eqref{PBth} and \eqref{boundary transformations} yield the falloff of the local parameters,
\begin{equation}
    \theta_u=\mathcal{O}(r^{-1})\,, \qquad  \theta=\mathcal{O}(r^0)\,, \qquad \eta=\mathcal{O}(r^0)\,.
\end{equation}
This shows that the transformations associated to $\theta$ and $\eta$ might lead to nontrivial transformations on the boundary. In the next subsection, we will explore the charges associated to them.

\subsection{Improved generators and charges}

We have two smeared generators \eqref{G,S}, but they are not differentiable, unless we add  the surfaces terms, or charges $Q[\theta ]$ and $Q_s[\theta ]$, and obtain the improved generators
\beq
G_{Q}[\theta ] =G[\theta ]+Q[\theta ]\,, \qquad
S_{Q}[\eta ] = S[\eta ]+Q_s[\eta ]\,. \label{improved}
\eeq
The precise form of these surface integrals can be found from the differentiability of $G_{Q}[\theta ]$ and $S_{Q}[\eta ]$, that is, asking that the variations
\begin{equation}
\begin{split}
\delta G_{Q}[\theta ]&=\int \diff^{3}x\,\left( \frac{G_{Q}[\theta ]}{\delta A_{\mu }}\,\delta A_{\mu }+\frac{G_{Q}[\theta ]}{\delta \pi ^{\mu }}\,\delta \pi ^{\mu }\right) \,, \\
\delta S_{Q}[\eta ] &=\int \diff^{3}x\,\left( \frac{S_{Q}[\eta ]}{\delta A_{\mu }}\,\delta A_{\mu }+\frac{S_{Q}[\eta ]}{\delta \pi ^{\mu }}\,\delta
\pi ^{\mu }\right) \,,
\end{split}
\end{equation}
do not possess any surface terms. It yields the expressions for the variational derivatives of $G_{Q}$ and $S_{Q}$,
 \beq
\begin{split}
\frac{\delta G_{Q}[\theta ]}{\delta \pi ^{i}} &=-\partial _{i}\theta \,,
\qquad \qquad \qquad \quad \frac{\delta G_{Q}[\theta ]}{\delta \pi ^{u}}=\theta _{u}\,, \\
\frac{\delta S_{Q}[\eta ]}{\delta A_{r}} &=-\frac{1}{e^2}\, \sqrt{\gamma }\nabla_{A}\tilde\eta^{A}\,,\qquad \quad \frac{\delta S_{Q}[\eta ]}{\delta \pi^{A}}=\epsilon\, \eta
_{A}\,,
\end{split}
\eeq
consistent with the PBs used to obtain \eqref{PBth} and \eqref{PBeta}, as well as the variations 
\begin{equation}
    \delta Q[\theta ] =-\int \diff^{3}x\, \partial _{r}\left( \theta \,\delta
\pi ^{r}\right) , \qquad \delta Q_{s}[\eta ] =
 \frac{1}{e^2} \int \diff ^{3}x\,
\partial _{r}\left( \sqrt{\gamma }\,\tilde\eta^{A}\delta A_{A}\right)  \,.
\end{equation}
Assuming that the parameters are field-independent, we can easily obtain the $\delta$-integrated form of the canonical charges, 
\beq
Q[\theta ]=-\oint \diff ^{2}y \,\theta\, \pi ^{r}\,,  \qquad
Q_{s}[\eta ] =\frac{1}{e^{2}} \oint \diff ^{2}y\,\sqrt{\gamma }\,\tilde\eta^{A}A_{A}\,, \label{charges}
\eeq
where the integral is performed over  the two-sphere  at the infinity or, more precisely, over a contour around the center of the complex plane. In that sense, the fields under the integral are $\pi^r_{(0)}$ and $A_{(0)A}$, but we will not write the index $(0)$, as it is clear that only these coefficients will contribute in the limit $r \to \infty$. The obtained charges are, in fact, the infinite number of global charges, corresponding to the Laurent expansion of the local parameters.  They are both the charges and the symmetry generators because, on the constraint surface, $G_Q[\theta]= Q[\theta]$ and $S_Q[\eta]= Q_s[\eta]$.

The conserved quantities \eqref{charges} can be compared to the ones known in the
literature in Lagrangian formalism after using the definition of
the radial momentum and the radial Maxwell's equation in the standard
falloff \eqref{standard}, namely $\nabla ^{A}A_{(0)A}=-\epsilon r^{2}F_{ur}+%
\mathcal{O}\left( r^{-1}\right) $. As a result, $Q[\theta ]$ acquires the
form of the known electric large $\mathrm{U}(1)$ gauge transformation \cite{He:2014cra}, while the second charge matches (up to a term of the form $Q[\theta ]$, with $\theta $ being a function of $\eta _{A}$) to the magnetic large gauge transformation
\cite{Campiglia:2016hvg}. In derivation, one should apply the Helmholtz-Hodge decomposition
of the vector $\eta_{A}$ to two scalars.\footnote{We thank an anonymous referee for bringing up this comparison to us.}

In the context of soft theorems associated to asymptotic symmetries in flat
spacetime, the electric charge $Q[\theta ]$ generates the usual Weinberg's
soft photon theorem, while the magnetic charge $Q_{s}[\theta ]$ generates the magnetic soft photon theorem \cite{Strominger:2015bla}.

\subsubsection*{Invariance of the symplectic form}
\label{symMax}
In Hamiltonian analysis, an essential role is played by the canonical symplectic $2$-form in the phase space, 
\beq
\label{symp}
\omega=\int \diff ^{3}x \, \delta \pi^\mu \wedge \delta A_\mu \,,
\eeq
where the differential $\delta$ acts  as an exterior derivative in this space. The result \eqref{symp} can be derived directly from $\omega$.

It is important to point out that precisely the improved generators \eqref{improved} are the ones that keep unchanged the symplectic form \eqref{symp}. Indeed, the transformations generated by Hamiltonian vector fields $X$ preserve the symplectic structure if and only if the Lie derivative $\mathscr{L}_X
\equiv \delta \,i_X + i_X \delta $ acting on the canonical symplectic form vanishes,
\beq
\mathscr{L}_X \omega=0 \quad \Leftrightarrow \quad \delta(i_X \omega)=0\,. \label{closed}
\eeq
Here, $i_X$ denotes the interior product (contraction) with $X$ and, in the last step, we used the fact that the symplectic form is closed, $\delta \omega=0$. In turn, \eqref{closed} implies that $i_X\omega$ is locally exact, $i_X\omega=\delta(\cdots)$.

In the present case, the infinitesimal transformations given in \eqref{PBth} and \eqref{PBeta} are described by the vector fields 
\beq
X_\theta=\int \diff^3x \left(\theta_u \,\frac{\delta}{\delta A_u} - \d_i \theta \, \frac{\delta}{\delta A_i}\right)\,, \quad X_\eta=\int \diff^3x \left(\frac{ \sqrt{\gamma }}{e^2}\,\,\nabla_{A}\tilde\eta^{A}\, \frac{\delta}{\delta \pi^r}+\epsilon\eta_A\,\frac{\delta}{\delta A_A}\right)\,.
\eeq
Proceeding as previously described, it is direct to show that the contraction of $X_\theta$ and $X_\eta$ with the symplectic form \eqref{symp} produces variations of the improved generators defined at the beginning of this section, 
\beq
i_{X_\theta} \omega=-\delta G_Q[\theta]\,, \qquad i_{X_\eta} \omega= -\delta S_Q[\eta]\,. \label{symplectic_gen}
\eeq
Since the right-hand sides of these equations are total variations, from \eqref{closed} they are equivalent to  $\mathscr{L}_{X_\theta} \omega=0$ and $\mathscr{L}_{X_\eta} \omega=0$. As a consequence, $X_\theta$ and $X_\eta$ leave the symplectic form invariant, and the improved  generators are  identified with the action of the corresponding vector field.

\subsection{Charge algebra}
\label{ChargeAlgebra}
  
Consider a set of local transformations with  parameters $\alpha$ and improved generators $G_Q[\alpha]=G[\alpha]+Q[\alpha]$.  Their PBs can be written in a simple way in terms of the symplectic form \eqref{symp} as
\beq
\label{br1}
 \{G_Q[\alpha],G_Q[\beta]\}\equiv i_{X_\alpha} i_{X_\beta} \omega \,,
\eeq
which is useful for the computation of an algebra of improved generators. First we notice that the exterior derivative acts on a PB of improved generators as an interior product of the symplectic form with respect to the Lie brackets $[X_\alpha,X_\beta]$,
\beq
\label{br2}
\delta\left(\{G_Q[\alpha],G_Q[\beta]\}\right)=-i_{[X_\alpha,X_\beta]} \omega\,.
\eeq
Then, because this Lie bracket is a Hamiltonian vector field itself, the r.h.s.~of the above expression can be written in terms of the improved generator as $\delta G_Q[[\alpha,\beta]]$, so that $G_Q[[\alpha,\beta]]$ is associated to the action of the vector field $[X_\alpha,X_\beta]$.
With all of that, the expression under the exterior derivative $\delta$ is determined only up to a constant $C[\alpha,\beta]$, leading to the projective representation of the algebra
\beq
\label{br3}
\{G_Q[\alpha],G_Q[\beta]\}=G_Q[[\alpha,\beta]]+C[\alpha,\beta]\,,
\eeq
where $C[\alpha,\beta]=-C[\beta,\alpha]$ is a central element, since it is independent of the canonical variables. 

Besides, the reduced phase space contains only physical fields, where all constraints are identically satisfied, implying $G[\theta ]=0$ and $G_{Q}[\theta ]=Q[\theta ]$. There, the PBs are replaced by Dirac brackets. From now on, we will work with Dirac brackets, but we will use the same notation for the sake of simplicity, because  two types of brackets cannot be confused. In reduced phase space, \eqref{br3} becomes an algebra of charges, 
\beq
\{Q[\alpha],Q[\beta]\}=Q[[\alpha,\beta]]+C[\alpha,\beta]\,. \label{charge algebra_def}
\eeq
A well-known example of a centrally extended classical  algebra corresponds to a realisation of asymptotic symmetries in AdS$_{3}$
spaces \cite{Brown:1986nw}.

In our case, the reduced phase space contains the fields $A_{(0)A}$ and $\pi _{(0)}^{r}$, and $Q[\theta ]$ and $Q_{s}[\eta ]$ are an infinite number of Abelian charges. To investigate an existence of a central extension, we use \eqref{br1} and find 
\beq
\{Q[\theta _{1}],Q[\theta _{2}]\}=0\,, \label{Q1Q1}
\eeq
concluding that this $\mathrm{U}(1)$ subalgebra does not possess a central extension, which is a known result \cite{Henneaux:2018gfi,Bunster:2018yjr}. As regards the generators $S_{Q}[\eta ]$, we get
\beq
\{Q_{s}[\eta _{1}],Q_{s}[\eta _{2}]\} =  \int \diff ^{3}x \, \left( \delta_{\eta_2}\pi^\mu \delta _{\eta
_{1}}A_{\mu}-\delta_{\eta_1}\pi^\mu \delta _{\eta
_{2}}A_{\mu}\right)  =0\,, \label{Q2Q2}
\eeq
also finding that it does not possess a central charge. Finally, we analyze the mixed algebra of the generators  $Q[\theta ]$ and $Q_s[\eta ]$, and get
\beq
\{Q[\theta ],Q_{s}[\eta ]\} =
\frac{1}{e^2} \oint \diff^{2}y \,\sqrt{\gamma }\,\tilde\eta^{A}\partial _{A}\theta \,. \label{Q1Q2}
\eeq
This is an Abelian brackets with a non-trivial central extension
\begin{equation}
C[\theta ,\eta ]=\frac{1}{e^2} \oint \diff^{2}y \,\sqrt{\gamma }\,\tilde\eta^{A}\partial
_{A}\theta \neq 0 \,.
\label{central}
\end{equation}
Due to the existence of the non-vanishing central term, the charges $Q[\theta ]$ and $Q_{s}[\eta ]$ can be seen as holographic conjugate pairs living on the celestial sphere. This has been discussed in \cite{Donnay:2018neh}, where $Q[\theta ]$ is associated to the conformally soft photon mode, while $Q_{s}[\eta]$ generates a Goldstone current. In order to make a more transparent connection with these operators and its relation with celestial conformal field theory,  we analyse the structure of the algebra \eqref{Q1Q1}--\eqref{central} in the Laurent modes on the two-sphere.

\subsubsection*{Mode expansion of the algebra}

Any tensor field $\psi(y)$ on the celestial sphere $y^A=(z,\bar{z})$ can be expanded in the Laurent series with the coefficients $\psi_{nm}$ $(n,m \in \mathbb{Z})$, 
\begin{equation}
\psi(y)=\sum_{n=-\infty }^{+\infty }\sum_{m=-\infty }^{+\infty} \frac{\psi_{nm}}{z^{n+h}\bar{z}^{m+\bar h}} \quad \Leftrightarrow \quad \psi_{nm}=\oint \frac{\diff ^{2}y}{(2\pi \mathrm{i})^{2}}\,z^{n+h-1}\bar{z}^{m+\bar{h}-1}\psi(y)\,,  \label{Laurent}
\end{equation}
where we used $\oint \frac{dz}{z^{n}}=2\pi \mathrm{i}\,\delta _{n1}$. The powers $(h,\bar h)$ depend on the spin of the tensor $\psi$. Using these conventions, let us expand the local parameters on the boundary in the Laurent  series on the sphere.
To fix $(h,\bar h)$, we take into account that $\theta$ is the Lorentz scalar so that the pair is $(0,0)$, while the spin of $A_A$ is one, yielding $(1,0)$ for $A_z$, and $(0,1)$ for $A_{\bar z}$. Hence, the expansions are
\beq
\theta (y)=\sum_{n,m}\frac{\theta _{nm}}{z^{n}\bar{z}^{m}}\,, \qquad  \eta _{z}(y) =\sum_{n,m}\frac{\eta _{nm}}{z^{n+1}\bar{z}^{m}}\,, \qquad \eta_{\bar{z}}(y)=\sum_{n,m}\frac{\bar{\eta}_{nm}}{z^{n}\bar{z}^{m+1}}\,.
\label{Laurent_parameters}
\eeq
Similarly, the fields on the reduced phase space expand as 
\beq
A_{z} =\sum_{n,m}\frac{A_{nm}}{z^{n+1}\bar{z}^{m}}\,,\quad A_{\bar{z}}=\sum_{n,m}\frac{\bar{A}_{nm}}{z^{n}\bar{z}^{m+1}}\,,\quad
\pi^{r} =\sum_{n,m}\frac{\pi _{nm}}{z^{n}\bar{z}^{m}}\,. \label{APi}
\eeq
The charges \eqref{charges} become
\beq
Q[\theta ] =\sum_{n,m}\theta _{nm}G_{nm}\,,  \qquad 
Q_{s}[\eta ] = \sum_{n,m}\left( \eta _{nm}\bar{S}_{nm}+\bar{\eta}_{nm}S_{nm}\right) \,, 
\label{Laurent_charges}
\eeq
where the Laurent expansion of the generators has the form
\begin{equation}
    G_{nm}=4\pi^2\,\pi _{1-n,1-m}\,,\quad 
    S_{nm} = -\frac{4\pi^2}{e^2}\,A_{-n,-m}\,, \quad
    \bar{S}_{nm}= -\frac{4\pi^2}{e^2}\,\bar{A}_{-n,-m}\,.
\end{equation}
The central extension \eqref{central}, on the other hand, has an expansion
\beq
C[\theta ,\eta ] =\kappa \sum_{n,m}\sum_{k,l}\theta _{nm}\left( m\eta _{kl}+n\bar{\eta}%
_{kl}\right) \delta _{n+k,0}\delta _{m+l,0}\,,
\eeq
where the constant $\kappa$ is the $\mathrm{U}(1)$ level of the current algebra,
\begin{equation}
\kappa=\frac{4\pi ^{2}}{e^2} \,.
\end{equation}
With these results at hand, the  non-vanishing  brackets of the charge algebra mode expansion read
\beq
\{G_{nm},S_{kl}\} =\kappa n\,\delta _{n+k,0}\delta _{m+l,0}\,,   \quad
\{G_{nm},\bar{S}_{kl}\} =\kappa m\,\delta _{n+k,0}\delta _{m+l,0}\,.
\label{KacMoody}
\eeq

Before we proceed, let us find the complex conjugation properties of the fields and generators. Using that $\pi^r$ and $A=A_A \diff y^A$ are real, as well as the parameters $\theta$ and $\eta=\eta_A \diff y^A$, we find
\beq
\begin{split}
\begin{array}[b]{llllll}
A_{nm}^{*} & =\bar{A}_{mn}\,,\qquad \medskip  & \theta _{nm}^{*} &
=\theta _{mn}\,, & G_{nm}^{*} & =G_{mn}\,, \\
\pi_{nm}^{*} & =\pi_{mn}\,, \medskip  & \eta _{nm}^{*} & =\bar{\eta}_{mn}\,,\qquad  & S_{nm}^{*} & =\bar{S}_{mn}\,. 
\end{array}
\label{*fields}
\end{split}
\eeq

Furthermore, we can separate the left and right current algebras in \eqref{KacMoody} by performing a  change of the basis $\{G_{nm},S_{nm},\bar S_{nm}\} \to \{R_{nm},J_{nm},\bar J_{nm}\}$, such that $J_{nm}$ and $\bar{J}_{nm}$ commute, and the new generators have the following conjugation properties
\begin{equation}
R_{nm}^{*}=R_{mn}\,,\qquad J_{nm}^{*}=\bar{J}_{mn}\,,\qquad \bar{J}
_{nm}^{*}=J_{mn}\,.
\end{equation}
The unique possibility satisfying these conditions, up to a field-independent scaling of $G$ and $S\pm \bar{S}$, is
\beq
\begin{split}
R_{nm} &=G_{nm}+\frac{1}{2}\,S_{nm}+\frac{1}{2}\,\bar{S}_{nm} \,,\\
J_{nm} &=G_{nm}+\frac{1}{2}\,S_{nm}-\frac{1}{2}\,\bar{S}_{nm}\,,   \\
\bar{J}_{nm} &=G_{nm}-\frac{1}{2}\,S_{nm}+\frac{1}{2}\,\bar{S}_{nm}\,. 
\end{split}
\label{LRcurrents}
\eeq
The algebra \eqref{KacMoody} in the new basis becomes 
\begin{eqnarray}
\left\{ J_{nm},\;J_{kl}\right\}  &=&\kappa (n-m)\, \delta_{n+k,0}\delta _{m+l,0}\,,  \notag \\
\left\{ \bar{J}_{nm},\,\bar{J}_{kl}\right\}  &=&-\kappa (n-m)\,\delta _{n+k,0}\delta _{m+l,0}\,, \notag \\
\left\{ R_{nm},\,J_{kl}\right\}  &=&\kappa n\,\delta _{n+k,0}\delta _{m+l,0}\,, \label{KM extended}
 \\
\left\{ R_{nm},\bar{J}_{kl}\right\}  &=&\kappa \,m\,\delta _{n+k,0}\delta_{m+l,0}\,, \notag\\
\left\{ R_{nm},R_{kl}\right\}  &=&\kappa \,(n+m)\,\delta _{n+k,0}\delta_{m+l,0}\,, \notag
\end{eqnarray}
where all other brackets are zero. 

The currents $J_{nm}$ span an algebra that is a generalization of an Abelian Kac-Moody (KM) algebra. From \eqref{KM extended}, it can be seen that the current $J_{n0}/J_{0n}$ is a generator of the standard Abelian KM algebra with the level $\kappa / -\kappa$. The fact that they commute suggests that they correspond to holomorphic and antiholomorphic currents.
Similarly, $\bar J_{nm}$ contain the generators of other two independent Abelian KM algebras, whose levels have the opposite signs. Remarkably,  the generators $R_{nm}$ extend previous algebras non-trivially, such that the currents $R_{nm}$ themselves close the third pair of Abelian KM algebras, with all the levels positive. 
Obtained algebras are not independent because of non-trivial brackets between the currents $R$ and $J$, $\bar J$. 

More explicitly, the Abelian KM algebras with the currents $\{ J_{n0},J_{0n},\bar{J}_{n0},\bar{J}_{0n},$  $R_{n0},R_{0n}\}$ possess the levels 
\begin{eqnarray}
\left\{ J_{n0},\;J_{m0}\right\}  &=&\kappa n\,\delta _{n+m,0}\,,\qquad \ \ \, 
\left\{ J_{0n},\;J_{0m}\right\} =-\kappa n\,\delta _{n+m,0}\,,  \notag \\
\left\{ \bar{J}_{n0},\,\bar{J}_{m0}\right\}  &=&-\kappa n\,\delta_{n+m,0}\,,\qquad \left\{ \bar{J}_{0n},\, \bar{J}_{0m}\right\} =\kappa n\,\delta _{n+m,0}\,, \label{KMcurrents}\\
\left\{ R_{n0},R_{m0}\right\}  &=&\kappa n\,\delta _{n+m,0}\,,\qquad \ \ \;
\left\{ R_{0n},R_{0m}\right\} =\kappa n\,\delta _{n+m,0}\,,  \notag
\end{eqnarray}
where the only non vanishing mixed brackets among them are
\begin{equation}
\left\{ R_{n0},J_{m0}\right\} =\kappa n\,\delta _{n+m,0}\,,\qquad  \quad  \left\{
R_{0n},\bar{J}_{0m}\right\} =\kappa n\,\delta _{n+m,0}\,.
\end{equation}

From the six KM currents, the  global Abelian algebra is spanned by the generators $J_{00}$, $\bar{J}_{00}$ and $R_{00}$.  We can  deduce how they act on the boundary fields $\pi_{(0)} ^{r}$ and $A_{(0)}=A_{(0)A}\diff y^A$ using the transformation laws \eqref{PBth} and \eqref{PBeta}. We find that the action of $G_{00}$ is trivial, which is consistent with the fact that the electromagnetic $\mathrm{U}(1)$ transforms only  charged particles. Thus, only two generators act non-trivially on the boundary as total differentials $\delta _{\eta _{00}}A_{(0)}=\frac{\diff \bar{z}}{\bar{z}}$ and $\delta _{\bar{\eta}_{00}}A_{(0)}=\frac{\diff z}{z}$,
describing therefore the translations in the $\bar z$ and $z$ directions, respectively.

Finally, we observe that, in a space of tensors defined on the Riemann sphere $\mathbb{S}^2$, which depends on two complex coordinates $y^{A}=(z,\bar{z})$, a KM algebra is naturally formed only from the generators that are proportional to holomorphic and anti-holomorphic tensors, namely, the ones with the first or second index equal to zero. The fact that the algebra \eqref{KM extended} is spanned by a generalization of these tensors to the ones with  two independent indices, shows that an effective boundary field theory has a much richer symmetry structure.

\subsection{Sugawara construction}
\label{SUGAWARA}

It is known that a  classical Virasoro algebra without a central charge can be obtained from a Kac-Moody one using the Sugawara construction \cite{Sugawara:1967rw}.
We will apply this technique to obtain the generators quadratic in the four KM currents $\left\{ J_{n0},J_{0n},\bar{J}_{n0}, \bar{J}_{0n}\right\} $. Taking care of the level of each current given in \eqref{KMcurrents}, we define the Virasoro generators 
\begin{eqnarray}
L_{n} &=&\frac{1}{2\kappa }\sum_{k}J_{k0}J_{n-k,0}\,,\qquad \ \ \,  \bar{L}_{n}=\frac{1}{2\kappa } \sum_{k}\bar{J}_{0k}\bar{J}_{0,n-k}\,,  \notag \\
\mathcal{L}_{n} &=&-\frac{1}{2\kappa } \sum_{k}J_{0k}J_{0,n-k}\,,\qquad
\mathcal{\bar{L}}_{n}=-\frac{1}{2\kappa } \sum_{k}\bar{J}_{k0}\bar{J}_{n-k,0}\,, \label{U(1) currents} 
\end{eqnarray}
such that it is satisfied $L^*_n=\bar L_n$ and $\mathcal{L}^*_n=\bar{\mathcal{L}}_n$.  More general $\mathrm{U}(1)$ currents with two indices can also be defined through a Sugawara construction, replacing the index $0$ by a fixed integer $s$. Nonetheless,  an algebra closes linearly in that case only when $s=0$. Besides, because the generators quadratic in $R_{n0}$ and  $R_{0n}$ do not commute  with other four currents, their inclusion in the algebra requires an addition of new Virasoro generators. We will leave them out of this discussion.
As a result, we find the following Virasoro algebras,
\begin{eqnarray}
\left\{ L_{n},L_{m}\right\}  &=&\left( n-m\right) \,L_{n+m}\,,\qquad \left\{
\bar{L}_{n},\bar{L}_{m}\right\} =\left( n-m\right) \,\bar{L}_{n+m}\,,  \notag
\\
\left\{ \mathcal{L}_{n},\mathcal{L}_{m}\right\}  &=&\left( n-m\right) \,\mathcal{L}_{n+m}\,,\qquad \left\{ \mathcal{\bar{L}}_{n},\mathcal{\bar{L}}_{m}\right\} =\left( n-m\right) \,\mathcal{\bar{L}}_{n+m}\,,
\label{Virasoro} 
\end{eqnarray}
where all other brackets vanish.
Their non-zero brackets with the currents are
\begin{equation}
\begin{split}
\left\{ L_{n},J_{m0}\right\} & =-mJ_{n+m,0}\,,\qquad \left\{ \bar{L}_{n},%
\bar{J}_{0m}\right\} =-m\bar{J}_{0,n+m}\,, \\
\left\{ \mathcal{L}_{n},J_{0m}\right\} & =mJ_{0,n+m}\,,\qquad \ \ \left\{
\mathcal{\bar{L}}_{n},\bar{J}_{m0}\right\} =m\bar{J}_{n+m,0}\,,
\end{split}
\end{equation}
and with the generators $R_{nm}$  
\begin{equation}
\left\{ R_{n0},L_{m}\right\} =n\,J_{n+m,0}\,,\qquad \left\{ R_{0n},\bar{L}_{m}\right\} =n\,\bar{J}_{0,n+m}\,,
\end{equation}%
while $\mathcal{L},\mathcal{\bar{L}}$ commute with $R$. 
For completeness,
using eq.~\eqref{KM extended}, we also add non-zero brackets
between the currents,
\begin{equation}
\left\{ R_{n0},J_{m0}\right\}  =\left\{
R_{0n},\bar{J}_{0m}\right\} =
\left\{ R_{n0},R_{m0}\right\} =\left\{ R_{0n},R_{0m}\right\}  =\kappa n\,\delta
_{n+m,0}\,.
\end{equation}

In that way, we found the extended algebra between the currents with one index zero and Virasoro generators ($L,\bar{L},\mathcal{L},\mathcal{\bar{L}}$).
Nonetheless, these are not the only possible Virasoro algebras that can be
constructed from the currents $\{R_{nm},J_{nm},\bar{J}_{nm}\}$. A
particular case of physical interest,  motivated by the charge  $Q[i_{Y}A_{(0)}]$ that makes the Poincaré generator differentiable (see eq.~\eqref{B} in Discussion), are bilinears constructed from $G$ and $S$,$\bar{S}$, 
\begin{equation}
K_{n}=\frac{1}{\kappa }\,\sum_{k}G_{k0}S_{n-k,0}\,,\qquad \bar{K}_{n}=\frac{1}{\kappa }\,\sum_{k}G_{0k}\bar{S}_{0,n-k}\,, \label{KK}
\end{equation}
where $K_{n}^{*}=\bar{K}_{n}$. It is direct to check that these generators also close two independent Virasoro algebras,
\begin{equation}
\left\{ K_{n},K_{m}\right\} =\left( n-m\right) K_{n+m}\,,\qquad \left\{ \bar{K}_{n},\bar{K}_{m}\right\} =\left( n-m\right) \bar{K}_{n+m}\,.
\end{equation}
Using the definitions of the currents \eqref{LRcurrents}, the generators 
can be expressed as
\begin{eqnarray}
K_{n} &=&\frac{1}{2\kappa }\,\sum_{k}\left( J_{k0}+\bar{J}_{k0}\right)
\left( R_{n-k,0}-J_{n-k,0}\right) \,,  \notag \\
\bar{K}_{n} &=&\frac{1}{2\kappa }\,\sum_{k}\left( J_{0k}+\bar{J}_{0k}\right)
\left( R_{0,n-k}-\bar{J}_{0,n-k}\right) \,, \label{K-barK}
\end{eqnarray}
showing that they are independent on $(L,\bar{L},\mathcal{L},\mathcal{\bar{L}})$. It is not possible to extend the algebra to include all six Virasoro generators at the same time, because it becomes nonlinear.  

At the end, it is worth noticing that the Sugawara construction in the framework of Hamiltonian formalism has been used to explore a dual dynamics of three dimensional asymptotically flat Einstein gravity at null infinity \cite{Salomonson:1989fw,Barnich:2013yka} and in asymptotically AdS space \cite{Banados:1994tn,Blagojevic:2009hb}. In our case, again, the fact that we have found more Virasoro generators than the ones appearing in a standard conformal field theory, suggests that an effective holographic theory that arises here has much richer dynamics and symmetries.

\section{Yang-Mills theory in the null frame}

The symmetry structure found in the canonical formulation of the electromagnetism on the light front is expected to exist also in Yang-Mills theory, which is particularly interesting because it is invariant under a non-Abelian gauge group $\mathcal{G}$ of dimension $d=\dim \mathcal{G}$. It is described by the action functional
\beq
I[A]=-\frac{1}{4 g^2} \int \diff ^4x \sqrt{\mathfrak{g}} \, F^{\mu \nu}_a F^a_{\mu \nu}\,,
\eeq
of a Lie-algebra valued gauge field $A_{\mu}=A^a_\mu T_{a}$,  where $g$ is the coupling constant. $T_a$ are $d$ anti-Hermitean generators of a compact non-Abelian semi-simple group $\mathcal{G}$, whose Lie algebra and a Cartan-Killing metric read\footnote{The Abelian group $\mathcal{G}=U(1)$, although not compact, is also included in this discussion when $d=1$.} 
\beq
[T_a,T_b]=f_{ ab}^{\;\;\,c}T_c\,, \qquad g_{ab}= -\mathrm{Tr}(T_a T_b)\,, \label{Lie algebra}
\eeq
with the structure constants of the Lie algebra $f_{ ab}^{\;\;\,c}$.  The Cartan-Killing metric  $g_{ab}$ and its inverse $g^{ab}$ raise and lower group indices. 
The non-Abelian field strength is given as usual by $F_{\mu \nu}^a=\d_\mu A^a_{\nu}-\d_\nu A^a_\mu+f^{a}_{\;\; bc} A^b_\mu A^c_{\nu}$. 

\subsection{Hamiltonian action and its symmetries}

Using the Minkowski background in the retarded (advanced) Bondi coordinates \eqref{flat} with $\epsilon^2=1$,  the canonical momenta are given by
\beq
\pi^\mu_a=\frac{\partial \mathcal{L}}{\partial \dot{A^a}_{\mu }}=- \frac{1}{g^{2}} \sqrt{\mathfrak{g}}\, F^{u \mu}_a\,,
\eeq
and the fundamental PBs read \beq
\{A^a_{\mu}(x),\pi^\nu_b(x')\}=\delta^{a}_{b}\delta^{\mu}_{\nu} \delta^{(3)}(x-x')\,.
\eeq
Proceeding similarly as in the Abelian case, the primary constraints are
\beq
\pi^u_a\approx0\,, \qquad \chi^{A}_a\equiv\epsilon\,\pi^A_a- \frac{1}{g^2}\, \sqrt{\gamma} \gamma^{AB} F_{rB}^{a}\approx0\,,
\eeq
and the secondary constraints, which are a non-Abelian generalization of the Gauss law, have the form
\beq
\chi_a \equiv\mathscr{D}_i \pi^i_a\approx0\,.
\eeq
The covariant derivative acts on the group indices as $\dD_\mu v^a=\d_\mu v^a +f^{a}_{\;\; bc} A^b_\mu v^c\equiv [\dD_\mu]^a_{\ c} v^c$, where for convenience we defined a differential operator
\begin{equation}
    [\dD_\mu]^a_{\ c}=\delta^a_c\d_\mu +f^{a}_{\;\; bc} A^b_\mu\,.
\end{equation}
The above constraints satisfy the following non-vanishing  PBs,\footnote{In the derivation, we used the Jacobi identity
$f_{ad}^{\;\;\; e}\,f_{bc}^{\;\; d}+f_{bd}^{\;\;\; e}\,f_{ca}^{\;\; d}+f_{cd}^{\;\;\; e}\,f_{ab}^{\;\; d}=0$.}
\begin{eqnarray}
\left\{ \chi _{a}(x)\,,\;\chi _{b}(x')\right\}  &=&f_{ab}^{\;\; c}\,\chi _{c}(x)\,\delta ^{(3)}(x-x')\,, \notag \\
\left\{ \chi _{a}(x),\,\chi _{b}^{A}(x')\right\}  &=& f_{ab}^{\;\;
c}\,\chi^A _c(x)\,\delta ^{(3)}(x-x') \,, \label{ConstraintPB}\\
\left\{ \chi _{a}^{A}(x),\chi _{b}^{B}(x')\right\}  &=&\Omega
_{ab}^{AB}(x,x')\,, \notag 
\end{eqnarray}
where the non-Abelian symplectic matrix reads
\begin{equation}
\Omega _{ab}^{AB}(x,x')=-\frac{2\epsilon}{g^2} \, \sqrt{\gamma} \gamma^{AB} [\dD_r]_{ab}\,
 \delta^{(3)}(x-x') \,.\label{SymplecticYM}
\end{equation}
From \eqref{ConstraintPB} we conclude that, similarly as in the case of electromagnetism,  $\pi^u_a$ and $\chi_a$ are first class constraints. The first line makes it evident that $\chi _{a}$ close an algebra of $\mathcal{G}$, while the second line shows that $\chi _{a}^{A}$ transforms as
a $\mathcal{G}$-vector. The third line has to be further analyzed because the class of $\chi_a^A$ depends on the symplectic form. 

Now we can write out the total Hamiltonian density that contains also the multipliers $\lambda^a_A$,
\beq
\cH_{T}=g^2\,\frac{ \pi ^{r}_a\pi ^{r\,a}}{2 r^{2}\sqrt{\gamma }}+ g^2 \,\frac{\tilde{\pi}_{A}^a\pi_a^{A}}{2\sqrt{\gamma }}+\frac{\sqrt{\gamma }}{4g^2 r^{2}}\, \tilde{F}^{AB}_aF^a_{AB}-A_u^a\chi_a+ \lambda^a_A \chi^A_a+ \lambda^a_u \pi^u_a .  \label{HTYM}
\eeq
Using the algebra \eqref{ConstraintPB}, it can be shown that  all the constraints remain weakly zero during their evolution without any new condition, except of   $\dot{\chi}^{A}_a\approx0$, which partially determines a radial dependence of the multiplier $\lambda^a_A$ through the equation
\beq
\dD_r \lambda^a_A=-\frac{\epsilon g^2}{2\sqrt{\gamma}}\, \dD_r \pi^a_A-  \frac{1}{2r^2\sqrt{\gamma}}\, \tilde{\dD}^B (\sqrt{\gamma}\,F^{a}_{AB})\,  +\frac{\epsilon g^2}{2r^2} \,\dD_A\left( \frac{\pi^{r\,a}}{\sqrt{\gamma}}\right)\,.
\eeq
This is a covariantization of eq.~\eqref{lambdaEq} to a non-Abelian case, where we do not use $\nabla_A$ here, not to confuse it with a $\mathcal{G}$-covariant derivative $\dD_\mu$. Thus, one has to be careful when rising and lowering the angular indices because $\dD_\mu \gamma_{AB} \neq 0$. Again, $\lambda^a_A =\Lambda^a_A+\bar{\lambda}^a_A$ is defined up to an  arbitrary function $\Lambda^a_A(y)$ that points out to an  existence of a first class constraint.A form of the determined part of the multiplier, $\bar{\lambda}^a_A$, can be computed as well, but we will not need its explicit expression.\medskip 

For completeness, the Hamilton's equations of motion are
\begin{equation}
\begin{array}{llll}
\dot{A}_{u}^{a} & =\lambda _{u}^{a}\,,\medskip  & \dot{\pi}^{u} & =0\,, \\
\dot{A}_r^a= & \dfrac{g^2\pi ^{ar}}{r^2\sqrt{\gamma }} +\dD_rA_u^a\,,\medskip  & \dot{\pi}_{a}^{r} & =-f_{ab}^{\ \ c}A_u^b\pi
_{c}^r+\dfrac{1}{g^{2}}\,\dD_A\left( \sqrt{\gamma }\,\tilde\lambda_a^A\right)
\,, \\
\dot{A}_{A}^{a} & =\dfrac{g^{2}\tilde{\pi}_{A}^{a}}{\sqrt{\gamma }}+\dD_{A}A_u^a+\epsilon \lambda _{A}^{a}\,,  & \dot{\pi}^{A} &
=-f_{ab}^{\ \ c}A_u^b\pi _{c}^{A}-\dfrac{1}{g^{2}r^{2}}\,\dD_{B}\left( \sqrt{\gamma }\tilde{F}_{a}^{AB}\right) -\dfrac{\sqrt{\gamma }}{g^{2}}
\,\dD_r\tilde\lambda_{a}^{A}.
\end{array}
\label{HYM}
\end{equation}
which is a non-Abelian generalization of eqs.~\eqref{dot A} and \eqref{dot P}.

\subsection{Asymptotic behavior of the fields}
\label{AsympYM}

As shown in \cite{Strominger:2013lka,He:2015zea}, standard asymptotic behaviour of the Yang-Mills fields is similar to the one with the $\mathrm{U}(1)$ symmetry, given by eq.~\eqref{standard}, namely,
\begin{equation}
\begin{array}{llllll}
A^a_{u} & =\mathcal{O}\left( r^{-1}\right) \,,\quad  & A^a_{r} & =\mathcal{O}\left( r^{-2}\right) \,,\quad  & A^a_{A} & =\mathcal{O}(r^{0})\,,\medskip  \\
\pi_a ^{u} & =0 \,, & \pi_a ^{r} & =\mathcal{O}\left( r^{0}\right) \,, & \pi_a ^{A}
& =\mathcal{O}\left( r^{-2}\right) \,. \label{standardYM}
\end{array}
\end{equation}
Using the Hamilton's equations \eqref{HYM}, the multipliers behave asymptotically as \begin{equation}
\lambda^a_u=\mathcal{O}\left(r^{-1}\right) \,,\qquad \Lambda^a_A=\mathcal{O}\left( r^0\right) \,,\qquad A_u^a =\mathcal{O}\left(r^{0}\right) \,,
\end{equation}
where $\lambda^a_A =\mathcal{O}\left(r^{-1}\right)$ and $\bar\lambda^a_A =\mathcal{O}\left(r^{-1}\right)$, consistently with \eqref{StandardMultipliers}.
These asymptotic conditions are preserved by the action of Poincaré group \cite{He:2020ifr}.  

\subsubsection*{Second class constraints and zero modes}

Having boundary behaviour of the fields, we can find a set of second class constraints with the symplectic form given by \eqref{SymplecticYM}. Again, this matrix contains zero modes $V^a_A(x)$, because
\beq
\int \diff ^3x'\, \Omega^{AB}_{ab}(x,x')V^a_A(x')=0\,\quad
\Rightarrow \quad \dD_rV^a_A=0\,.
\eeq
Covariantly constant Lie-algebra valued vectors, $\dD_rV_A=0$, satisfy the differential equation $ \partial _{r}V_{A}=-[A_{r},V_{A}]$ that can be formally solved using an ordered exponential and the boundary condition $V_A(r \to \infty)=V_{(0)A}(y)$, 
\beq
V_A(x)=\dU^{-1} V_{(0)A}(y) \dU\,,\quad \dU=\dP \exp\left(-\int^\infty_r \diff r\; A_r \right)\,.
\eeq
If we assume the asymptotic behavior of $A_{r}$ as in \eqref{standardYM}, then the $r$-independent function $V_{(0)A}(x)$ remains arbitrary in the expansion of $V_A$,
\begin{equation}
V^a_A=V^a_{(0)A}+\frac{1}{r}\,f_{bc}^{\; \; a}\,V^b_{(0)A} A_{(2)r}^c+\mathcal{O}(r^{-2})\,,
\end{equation}
corresponding to a zero mode of the symplectic form.

\subsection{Conserved charges and their algebra}

Smeared generators associated to non-Abelian first class constraints have the form 
\beq
\begin{split}
G_Q[\theta ] &=\int \diff^{3}x\,\left( \theta^a \,\dD_{i}\pi_a^{i}+\theta^a
_u\,\pi_a ^{u}\right)+Q[\theta] \,, \\
S_Q[\eta ] &=\int \diff^{3}x \, \eta^a_{A} \chi^{A}_a +Q_s[\eta] \,,   \label{GSYM}
\end{split}
\eeq
where the following boundary terms, or Hamiltonian charges, are added to ensure differentiability of these generators,
\begin{equation}
Q[\theta ] =-\oint \diff^2 y\, \theta^a \pi^r_a \,, \qquad
Q_s[\eta ] =\frac{1}{g^2}\oint \diff^2 y\,\sqrt{\gamma}\,\eta^a_A  A_a^A\,.\label{QYM}  
\end{equation}
We used the same method as for the Abelian charges \eqref{charges}, and
assumed that the parameters are independent on the fields.

The generators \eqref{GSYM} produce the following transformations of the canonical fields, 
\begin{equation}
\begin{array}{llll}
\delta _{\theta ,\eta }A_{u}^{a} & =\theta_{u}^{a} \,, \medskip  & \delta_{\theta ,\eta }\pi _{a}^{u} & =0\,, \\
\delta _{\theta ,\eta }A_{r}^{a} & =-\dD_{r}\theta ^{a}\,,\medskip  & \delta_{\theta ,\eta }\pi _{a}^{r} & =f_{ab}^{\;\; c}\,\theta ^{b}\pi_{c}^{r}+\dfrac{1}{g^2}\,\dD_{A}\left( \sqrt{\gamma }\,\tilde\eta _{a}^{A}\right) \,, \\
\delta _{\theta ,\eta }A_{A}^{a} & =-\dD_{A}\theta ^{a}+\epsilon\, \eta_{A}^{a}\,,\quad  & \delta _{\theta ,\eta }\pi _{a}^{A} & =f_{ab}^{\; \; c}\,\theta ^{b}\pi _{c}^{A}-\dfrac{1}{g^2}\,\sqrt{\gamma }f_{ab}^{\;\; c}\,\tilde\eta_{c}^{A}A_{r}^{b}\,.
\end{array}
\end{equation}
These are the usual non-Abelian gauge transformations with the local parameters ($\theta^a_u,\theta^a$) that change the gauge fields as a covariant derivative, and internally rotate the components of the momenta (that was zero in the Abelian case). Additionally, there are new global transformations with the parameter $\eta_A^a$,
analogous to the magnetic large gauge transformations obtained in the Abelian case, which act only at the null boundary of the spacetime, when $r\to\infty$.

It is worth noticing that the fast falloff of the component $A_r$, as $1/r^2$ (see \eqref{standardYM}), suppresses the contributions of the $f_{ab}^{\;\; c}\,\eta_{Ac}$-terms in the transformation laws of the  momenta $\pi^i_a$ . In fact, the symmetry parameters behave as
\begin{equation}
    \theta_u^a=\mathcal{O}(r^{-1})\,,\qquad  \theta^a=\mathcal{O}(r^0)\,,\qquad 
    \eta^a_A=\mathcal{O}(r^0)\,.
\end{equation}
The parameters and constraints of the Yang-Mills theory are also summarized as in \eqref{summary}. Now it is straightforward to show, using the method outlined at the beginning of Subsection \ref{ChargeAlgebra},  that the charge algebra in the Yang-Mills theory acquires the form \eqref{charge algebra_def},
\begin{equation}
\begin{array}{ccl}
\{Q[\theta _{1}],\;Q[\theta _{2}]\} & = & Q[[\theta _{1},\theta _{2}]]\,,\medskip \\
\{Q[\theta ],\;Q_{s}[\eta ]\} & = & Q_{s}[[\theta ,\eta ]]+C[\theta ,\eta ]\,,\medskip
\\
\{Q_{s}[\eta _{1}],Q_{s}[\eta _{2}]\} & = & 0\,,
\end{array}
\label{algebraYM}
\end{equation}
where the commutator between the Lie-algebra valued parameters is $[\alpha,\beta]^c=f_{ab}^{\;\; c}\,\alpha^a\beta^b$. The charges $Q$ and $Q_s$ satisfy algebras associated to the gauge groups $\mathcal{G}$ and $\mathrm{U}(1)$, respectively. These two subalgebras are extended by the mixed brackets with the central extension
\begin{equation}
    C[\theta,\eta]= \frac{1}{g^2} \oint \diff^2y \, \sqrt{\gamma}\, \tilde\eta^A_a \partial_A \theta^a\,.
\end{equation}

\subsubsection*{Mode expansion}

We expand all the fields and parameters in Laurent series on the Riemann sphere similarly  as before, using the formula \eqref{Laurent}, adding only a group index.  The charges in modes become
\begin{equation}
G_{nm}^{a}=4\pi^2\pi _{1-n,1-m}^{a}\,,\qquad S_{nm}^{a}=-\frac{4\pi^2}{g^{2}}
\,A_{-n,-m}^{a}\,,\qquad \bar{S}_{nm}^{a}=-\frac{4\pi^2}{g^{2}}\,\bar{A}_{-n,-m}^{a}\,.
\end{equation}
The charge algebra \eqref{algebraYM} in modes is a non-Abelian generalization of the current algebra \eqref{KacMoody}, with non-vanishing brackets
\begin{eqnarray}
\left\{ G_{nm}^{a},G_{kl}^{b}\right\}
&=&f_{\;\;\;c}^{ab}\,G_{n+k,m+l}^{c}\,,  \notag \\
\left\{ G_{nm}^{a},S_{kl}^{b}\right\}
&=&f_{\;\;\;c}^{ab}\,S_{n+k,m+l}^{c}+\kappa ng^{ab}\delta _{n+k,0}\delta
_{m+l,0}\,,  \label{KMlike} \\
\left\{ G_{nm}^{a},\bar{S}_{kl}^{b}\right\}  &=&f_{\;\;\;c}^{ab}\,\bar{S}%
_{n+k,m+l}^{c}+\kappa mg^{ab}\delta _{n+k,0}\delta _{m+l,0}\,,  \notag
\end{eqnarray}
where the level is
\begin{equation}
    \kappa =\frac{4\pi^2}{g^2}\,. 
\end{equation}
This is an affine extension of 
the Lie algebra spanned by $d$ non-Abelian
generators $G_{00}^{a}$ of the group $\mathcal{G}$, and $2d$ Abelian  generators $S_{00}^{a}$, $\bar{S}_{00}^{a}$ transforming as vectors under the action of $\mathcal{G}$.

Note that the $\cN=4$ supersymmetric extension of Yang-Mills theory with $N$ colors can be seen as a holographic field theory dual to AdS gravity  \cite{Maldacena:1997re}. In that case,   the large $N$ limit has to be taken, keeping  the 't~Hooft coupling $\lambda =Ng^{2}$ finite \cite{tHooft:1973alw}. Applying the limit in our case,\footnote{We thank Francisco Rojas for pointing out this limit to us.} with $\mathcal{G}=\mathrm{SU}(N)$ and $\kappa=\frac{4\pi^2N}{\lambda}$, after the rescaling of the generators
$S\rightarrow \frac{1}{\sqrt{N}}\,S$ and $G\rightarrow \frac{1}{\sqrt{N}}\,G$ in the charge algebra \eqref{KMlike}, it becomes Abelian \eqref{KacMoody}, with the central extension $\kappa'=\frac{4\pi^2 }{\lambda }$ that remains finite.

Returning to a general $\mathcal{G}$, contrary to the Abelian case, the basis $(R_{nm}^{a},J_{nm}^{a},\bar{J}_{nm}^{a})$, defined in eq.~\eqref{LRcurrents}, does not decouple the currents $J_{nm}^{a}$ and $\bar{J}_{nm}^{a}$, because
\begin{equation}
 \left\{ J_{nm}^{a},\bar{J}_{kl}^{b}\right\} =\frac{1}{2}\,f_{\;\;\;c}^{ab}\,(J_{n+k,m+l}^{c}+\bar{J}_{n+k,m+l}^{c}) \neq 0\,.   
\end{equation}
The rest of the algebra in the new basis also acquires a cumbersome form.

Thus, we will perform a Sugawara construction using the bilinears \eqref{KK}, whose non-Abelian generalization is  
\begin{equation}
K_{n}=\frac{1}{\kappa }\,g_{ab}\sum_{k}G_{k0}^{a}S_{n-k,0}^{b}\,,\qquad \bar{K}_{n}=\frac{1}{\kappa }\,g_{ab}\sum_{k}G_{0k}^{a}\bar{S}_{0,n-k}^{b}\,,
\end{equation}
where $K_{n}^{\ast }=\bar{K}_{n}$. Then the commutators become
\begin{eqnarray}
\left\{ K_{n},K_{m}\right\}  &=&\frac{1}{\kappa ^{2}}\,f_{abc}\sum_{k,l}G_{k+l,0}^{c}S_{m-l,0}^{b}S_{n-k,0}^{a}  \notag \\
&&+\frac{1}{\kappa ^{2}}\,\sum_{k,l}\left( f_{abc}S_{k+m-l,0}^{c}+\kappa
kg_{ab}\delta _{k+m-l,0}\right) G_{l0}^{b}S_{n-k,0}^{a}  \notag \\
&&-\frac{1}{\kappa ^{2}}\,\sum_{k,l}\left( f_{abc}S_{l+n-k,0}^{c}+\kappa
lg_{ab}\delta _{l+n-k,0}\right) G_{k0}^{b}S_{m-l,0}^{a}\,,
\end{eqnarray}
and similarly for the complexly conjugated generators. The cubic terms vanish due to the symmetry arguments.
Namely, each $GSS$ term is proportional to
\begin{equation}
f_{abc}\sum_{k,l}G_{n+m-k-l,0}^{a}S_{k0}^{b}S_{l0}^{c}=0\,,
\end{equation}%
where the zero appears due to a contraction of a symmetric and antisymmetric
tensor in the indices $k,l$. Therefore, the only contribution to the algebra comes from the KM level term, leading to the classical Virasoro algebras
\begin{equation}
\{ K_{n},K_{m}\} =\left( n-m\right) K_{n+m}\,,\qquad \{ \bar{K}_{n},\bar{K}_{m}\} =\left( n-m\right) \bar{K}_{n+m}\,.
\end{equation}
However, although we have got two algebras, using the symmetry arguments is not enough to cancel out the cubic terms in the bracket $\{ K_{n},\bar{K}_{m}\} $, implying that the algebra spanned by $(K,\bar{K})$ and the currents is nonlinear. Namely, we get
\begin{equation}
\left\{ K_{n},\bar{K}_{m}\right\} =\frac{1}{\kappa ^{2}}\,f_{abc}\sum_{k,l}
\left(  G_{kl}^{a}S_{n-k,0}^{b}\bar{S}_{0,m-l}^{c}-G_{k0}^{a}S_{n-k,l}^{b}\bar{S}_{0,m-l}^{c} +G_{0,m-l}^{a}S_{n-k,l}^{b}S_{k0}^{c}\right) \,.
\end{equation}

We can, therefore, have either $K_{n}$ or  $\bar{K}_{n}$, but not both of them at the same time, if we restrict to linear algebras.
Impossibility to have both left and right Virasoro algebras at the same time reminds of a two-dimensional system of non-Abelian chiral bosons \cite{Salomonson:1988mk}. But, in our case, it does not mean yet that an effective field theory on the celestial sphere is chiral, because we have more generators at our disposal to construct a more general algebra.

The above examples show that an effective two-dimensional holographic field theory dual to a Yang-Mills model also possesses a rich and interesting dynamics and symmetries, worth of exploring.

\section{Discussion}

We construct a canonical representation of symmetry generators in electromagnetism and Yang-Mills theory, in the null foliation of Minkowski spacetime. We find the usual gauge transformations described by the gauge
group $\mathcal{G}$ in Hamiltonian formulation. At the null boundary,  we also find, apart from the expected large gauge
transformations previously mentioned, the Abelian asymptotic symmetries given by an infinite number of global parameters that transform as 1-forms on the celestial sphere. The total asymptotic symmetry is given by an affine extension of the algebra associated to the Lie group $\mathcal{G}\times \mathrm{U}(1)$, where the Abelian parameters are 1-forms $\eta =\eta _{A}\,\diff y^{A}$. A canonical representation of the classical charge algebra is described by
a centrally extended Kac-Moody algebra, with the level $\kappa $ proportional to the coupling constant of the theory, $1/e^{2}$ for electromagnetism and $1/g^{2}$ for the Yang-Mills theory. Recently, central extensions were found in a realization of an algebra of soft generators associated with gauge theories coupled to matter fields \cite{Melton:2022fsf}.

The additional asymptotic symmetries come from the zero modes contained in the new primary second class constraints, which typically appear in canonical analysis  in the null foliation. With this constraint, a na\"{\i}ve count of bulk degrees of freedom, given by the Dirac formula as $\mathrm{d.o.f.}=d-N_{I}-\frac{1}{2}N_{II}$, where  $d=\dim \mathcal{G}$ and $N_{I}$ ($N_{II}$) is a total number of first (second) class constraint, leads to the result $\mathrm{d.o.f.}=d$. This is not in agreement with the known fact that a gauge field has $\mathrm{d.o.f.}=2d$. For instance, an electromagnetic wave has two degrees of freedom, or two states of polarization. A physical reason for this discrepancy is that a system in a reference frame that is moving by the speed of light does not have access to all degrees of freedom. Namely, a half of them is contained in a Hamiltonian multiplier \cite{Alexandrov:2014rta}, and therefore not recognized as a propagating mode. In turn, the Dirac formula is not applicable when the corresponding Hamiltonian multiplier is only partially determined  \cite{Henneaux:1992ig}. This is exactly our case, since the symplectic matrix of second class constraint has zero modes responsible for the undetermined part $\Lambda_A(y)$ of the otherwise determined multiplier $\lambda_A(x)$.

On the other hand, an advantage of  studying the symmetries that leave invariant the symplectic form is that we can treat on equal footing all the underlying symmetries of the theory. In particular, invariance under the Poincaré group is spanned by the vector fields $\xi=\xi^{\mu}\d_\mu$ that are Killing directions in the flat space, $\dL_\xi \mathfrak{g}_{\mu \nu}=0$. 
For the Minkowski metric in Bondi coordinates, the Killing vectors are computed in \cite{Barnich:2010eb}.

To be more specific, let us consider the electromagnetic theory. A canonical representation of the generator of (global) Poincaré transformations in the light front can be found using the energy-momentum tensor  $T_{\;\;\nu }^{\mu }=\frac{1}{e^{2}}\left( F^{\mu \alpha }F_{\nu \alpha }-\frac{1}{4}\,\delta _{\nu }^{\mu }\,F^{\alpha \beta }F_{\alpha \beta }\right)$.
Then, the canonical variables  are introduced in $T_{\;\;\nu}^{\mu}$ using the momenta definition and the primary constraints \eqref{primary}, such that the energy-momentum tensor has components 
\begin{eqnarray}
\sqrt{\mathfrak{g}}\,T_{\;\;u}^{u} &=&-\frac{e^{2}}{2r^{2}\sqrt{\gamma }}\,(\pi ^{r})^{2}-\frac{e^{2}}{2\sqrt{\gamma }}\,\pi ^{A}\tilde{\pi}_{A}-\frac{\sqrt{\gamma }}{4r^{2}e^{2}}\,\tilde{F}^{AB}F_{AB}+\frac{\epsilon e^2}{\sqrt{\gamma}}\,\pi^A\chi _{A}\,,  \notag \\
\sqrt{\mathfrak{g}}\,T_{\;\;i}^{u} &=&-F_{ij}\pi ^{j}\,.
\end{eqnarray}
The component $T_{\;\;u}^{u}$ is weakly equal to the minus total Hamiltonian but, because we have  freedom to add constraints to the generators, we will make this equality strong. 
Thus, the energy-density generator $\mathscr{H}_u$ and the Poynting vector generator $\mathscr{H}_{i}$ are defined by $\mathscr{H}_u =\mathcal{H}_T \approx -\sqrt{\mathfrak{g}}\,T_{\;\;u}^{u}\,,$ and $
\mathscr{H}_{i} = F_{ij}\,\pi ^{j}= -\sqrt{\mathfrak{g}}\,T_{\;\;i}^{u}$.
The Poincaré transformations are generated by the functional
\begin{equation}
\mathscr{G}[\xi]=\int \diff ^{3}x\,\left( \xi^\mu \mathscr{H}_{\mu}-i_{\xi^i
}A\, \chi\right) \,, \label{P_generator}
\end{equation}
 with the global parameters $\xi ^{\mu}$ that are Poincaré Killing vectors.  It acts on a function of canonical variables  as $\delta _{\xi}\psi=\{\xi,\mathscr{G}[\xi]\}$, leading to 
\begin{eqnarray}
\delta _{\xi}A_u &= &\xi^u\lambda_u\,, \qquad \delta _{\xi}\pi^u \approx 0\,, \notag \\
\delta _{\xi}A_{r} &\approx &\frac{e^{2}}{r^{2}\sqrt{\gamma }}\,\xi^u\pi ^{r}+\xi ^{i}F_{ir}+\partial _{r}\left( \xi^{\mu }A_{\mu }\right) \,, \notag \\
\delta _{_{\xi}}A_{A} &\approx&\frac{e^{2}}{\sqrt{\gamma }}\,\xi^u\tilde{\pi}_{A}+\xi^{i}F_{iA}+\partial _{A}\left( \xi ^{\mu }A_{\mu }\right) +\epsilon\, \xi^u \lambda _{A}\,,
\label{PoincareVar} \\
\delta _{_{\xi}}\pi ^{r} &\approx&\partial _{i}\left( \xi^{i}\pi ^{r}\right) -\partial _{i}\xi^{r}\pi ^{i}+\frac{\sqrt{\gamma }}{e^{2}}\,\nabla_{A}\left( \xi^u\tilde\lambda^{A}\right)\,, \notag\\
\delta _{_{\xi}}\pi ^{A} &\approx&-\frac{1}{r^{2}e^{2}}\,\partial _{B}\left( \sqrt{\gamma }\,\xi^u \tilde{F}^{AB}\right) +\partial _{i}\left( \xi^{i}\pi ^{A}\right) -\partial _{i}\xi^{A}\pi ^{i}-\frac{\sqrt{\gamma }}{e^{2}}\,\xi^u\partial _{r} \tilde\lambda ^{A} \,.  \notag
\end{eqnarray}
Similarly as for gauge generators, differentiability of the Poincaré generator is restored
by addition of a suitable 
boundary term that assumes only $\chi ^{A}\approx 0$ and $\delta \Lambda_A=0$, given by 
\begin{equation}
B[\xi ]\approx-Q\left[
i_{\xi}A_{(0)}\right] -Q_{s}\left[ \xi^u \Lambda \right] \,, \label{B}
\end{equation}
where the charges have the form \eqref{charges}. Remarkably, this result can be recovered
 directly from the symplectic form, yielding
$i_{X_\xi} \omega \approx -\delta(\mathscr{G}[\xi] +B[\xi])$, which implies that Poincaré symmetry is indeed canonically realized. In addition, the transformation law  \eqref{PoincareVar} is used to prove that the chosen boundary conditions are invariant under Poincaré transformations.  It is straightforward to repeat the same computation in the Yang-Mills case. The above transformation laws and their non-Abelian generalization are also used to prove that the boundary conditions \eqref{standard} and \eqref{standardYM} are invariant under Poincaré transformations.

Looking at the obtained results from another perspective, let us recall that the asymptotic gauge parameter 1-form $\eta $  transforms only the null boundary fields $A_{(0)}=A_{(0)A}\,\diff y^{A}$ and $\pi_{(0)}^r$. At the same time, these fields  play an essential role in a holographic quantum field theory, with the vacuum state $\left\vert A_{(0)}\right\rangle $. Then the improper transformations generated by $\mathrm{e}^{Q_{s}[\eta ]}$ do not keep this state invariant, but they change it to $\mathrm{e}^{Q_{s}[\eta ]}\left\vert
A_{(0)}\right\rangle =\left\vert A_{(0)}+\eta \right\rangle $. It leads  to an
interpretation of $A_{(0)}$ as a holographic Goldstone
mode \cite{Strominger:2017zoo}.

One can arrive at the similar conclusion using a different approach. Namely, asymptotic symmetries obtained here have an interesting mathematical  structure of a 2-group. Specifically, 2-groups emerge in a theory invariant under both 0-form and 1-form symmetries  \cite{Cordova:2018cvg}, such as in our case $ \mathcal{G}\times\mathrm{U}(1)$,
where the Abelian group corresponds to a 1-form symmetry. The concept of 2-groups \cite{Baez:2004hg}
and 2-algebras \cite{Baez:2004sg} have become an important area of research in mathematics and physics, with applications in  topological quantum field theory and algebraic topology, among others.  In electromagnetism, for instance, where the 0-form symmetry is $\mathcal{G}=\mathrm{U}(1)
$,  it has been shown that Maxwell electrodynamics realizes a 1-form symmetry in a CFT$_{4}$ \cite{Gaiotto:2014kfa,Hofman:2018lfz}. Furthermore, 
Goldstone photons are associated with the 2-group structure of the theory. It was also shown that the 1-form charges obey an algebra with central extension
 in \cite{Hofman:2018lfz}.
 
In this line of discussion, the next natural step is to apply obtained results to a holographic field theory arising on the null boundary of gauge theories. This would allow us to calculate the properties of an extended version of a CFT found here, by studying the features of its underlying algebraic structure. Another question of interest that emerges in a similar context  is related to asymptotic symmetries on the light front of $D$-dimensional gauge theories. The last but not the least open question would be to describe
the symmetries on the light front of an asymptotically flat
gravitational theory.

Let us also emphasize that we were dealing in this text with classical algebras only. Including the quantum effects of normal ordering would modify the algebra, such as add a central extension in the Virasoro case.

\section*{Acknowledgments}

We thank Glenn Barnich, Milutin Blagojevi\' c, Jordan Francois, Oscar Fuentealba, Marc Henneaux, Pantelis Panopoulos, Alfredo P\'erez, Arash Ranjbar, Francisco Rojas, Dejan Simi\'c and Ricardo Troncoso for useful discussions. This research has been supported by FONDECYT grants 11190427, 1190533, 1210635, 1211545, 1221920, 1230853 and 1231779. We would also like to acknowledge the support of ANID/ACT210100 ANILLO grant.

\bibliography{bibAsym}
\bibliographystyle{utphys}
\end{document}